\documentclass[11pt,a4paper]{article}

\usepackage[a4paper,margin=1in]{geometry}
\usepackage[utf8]{inputenc}
\usepackage[T1]{fontenc}
\usepackage{lmodern}
\usepackage{microtype}
\usepackage{amsmath,amssymb,mathtools,bm}
\usepackage{graphicx}
\usepackage{xcolor}
\usepackage{booktabs}
\usepackage[numbers,sort&compress]{natbib}
\usepackage[hypertexnames=false]{hyperref}

\hypersetup{
  colorlinks=true,
  linkcolor=blue,
  citecolor=blue,
  urlcolor=blue,
  pdftitle={Orbital Nodal Phase as a Pipeline Invariant in Black Hole Timing},
  pdfauthor={Mehmet Baran Okten}
}

\title{Orbital Nodal Phase as a Pipeline Invariant in Black Hole Timing}

\author{
Mehmet Baran \"{O}kten\\
Faculty of Engineering and Natural Sciences, Sabanc{\i} University\\
Istanbul, 34956, T\"{u}rkiye\\
\texttt{baran.okten@sabanciuniv.edu}
}

\date{}

\begin{document}

\maketitle

\begin{center}
\small
Accepted for publication in \emph{International Journal of Modern Physics D}. This is the author accepted manuscript.
\end{center}

\begin{abstract}
Timing analyses of accreting black holes often package nodal information in ways that depend on benign choices of time and azimuthal convention. We identify the corresponding pipeline-invariant content for slightly tilted circular rings and express it as an orbital nodal phase, $\Delta\psi_{\rm orb}$. In Kerr, this quantity gives the clean geodesic baseline for nodal timing: it equals the nodal precession per orbit, is invariant under the benign remappings considered here, and, for prograde Kerr spin, decreases monotonically with radius outside the innermost stable circular orbit. A fixed-$\Omega_\phi$ transport framework then isolates genuine metric sensitivity from trivial radius drift and provides the natural framework for far-field quadrupolar and higher-multipolar timing-response calculations. Two small analysis-level effects are also identified, namely a second-order bias from coherent radial breathing and the absence of an intrinsic geometric offset from exact slow fixed-$\Omega_\phi$ loops. A limited published-data illustration for GRO J1655$-$40 shows that the observational proxy for $\Delta\psi_{\rm orb}$ can be reconstructed directly from standard reported quasi-periodic oscillation frequencies once an orbital-frequency anchor and an identification convention are specified. Within the thin-ring limit, $\Delta\psi_{\rm orb}$ therefore provides a pipeline-robust reporting quantity and a Kerr-baseline diagnostic for source-level, simulation-level, and strong-gravity comparison applications.
\end{abstract}


\section{Introduction}\label{sec:intro}

The orbital and epicyclic frequencies defined with respect to Boyer--Lindquist (BL) time in the Kerr geometry form the backbone of strong-field timing models for accreting black holes, from the classic derivations of geodesic motion to modern interpretations of quasi-periodic oscillations \cite{BPT72,Chandra,MTW,StellaVietri1998,IngramMotta2020}. Within this framework, nodal precession of slightly tilted rings is usually discussed through the familiar ratio $\Omega_\theta/\Omega_\phi$, whose behavior outside the innermost stable circular orbit (ISCO) reflects the interplay of frame dragging and quadrupolar curvature. In practical pipelines, however, recorded time series often undergo benign redefinitions, such as mixing of time with azimuth along a ring or relabelling of the azimuthal coordinate, that complicate cross-epoch and cross-instrument comparisons even when the underlying physics is unchanged \cite{BPT72,StellaVietri1998}.

Recent developments make a pipeline-invariant nodal diagnostic timely. Strong-gravity tests from the Galactic Center and horizon-scale imaging increasingly constrain packages of multipoles and light-propagation systematics that are complementary to timing: the GRAVITY Collaboration has measured Schwarzschild precession for S2 around Sgr A* \cite{GRAVITY2020}, and the Event Horizon Telescope has extracted Kerr-consistent shadows and performed null-hypothesis tests with M87* \cite{EHT2019}. In parallel, gravitational-wave consistency tests now probe the Kerr hypothesis in highly dynamical, strong-field mergers \cite{LIGOtests2021}. Timing sits between these regimes and follows near-equatorial dynamics at radii set by the observed $\Omega_\phi$, while remaining directly sensitive to frame dragging and the quadrupole.

The contribution of this paper has two parts. First, for slightly tilted circular rings, the nodal timing content invariant under the benign time and azimuth redefinitions relevant to timing pipelines is the nodal phase accumulated per orbital cycle,
\begin{equation}
\Delta\psi_{\rm orb}
=
2\pi
\left(
1-\frac{\Omega_\theta}{\Omega_\phi}
\right).
\end{equation}
Second, when departures from the Kerr baseline are compared at fixed observed orbital frequency, the appropriate comparison geometry is the fixed-$\Omega_\phi$ transport viewpoint rather than the fixed-$r$ viewpoint. This separates genuine metric sensitivity from trivial radius relabelling and provides the natural framework for far-field quadrupolar and higher-multipolar timing-response calculations.

For a single idealised orbit, $\Delta\psi_{\rm orb}$ packages the nodal timing information into the scalar most directly suited to cross-pipeline comparison and transport-based calculations at fixed $\Omega_\phi$. In observational applications adopting the relativistic precession model, the corresponding proxy is simply $\Delta\psi_{\rm orb}/(2\pi)=\nu_{\rm nod}/\nu_{\rm HF}$. Beyond this invariant packaging, the framework yields two analysis-level results with direct interpretive value. Coherent radial breathing produces a second-order bias in the mean nodal phase, while exact slow evolution in a two-parameter control space produces no intrinsic geometric offset under fixed-$\Omega_\phi$ transport. A published-data benchmark for GRO J1655$-$40 then shows that the same quantity can be reconstructed directly from standard reported QPO frequencies once the orbital-frequency anchor and identification convention are stated. For both corrections we provide leading-order scalings and conservative magnitudes anchored in the standard epicyclic formulae \cite{Torok2005,Kato2001}.

The geometric language is not introduced because observational pipelines must operate with differential forms. Operationally, one still reconstructs the same scalar from the measured frequency pair once an orbital frequency anchor and an identification convention have been fixed. The point of the present formulation is instead to identify explicitly which combination of the standard timing frequencies is invariant under the admissible time and azimuth redefinitions considered here, to prove that invariance directly, and to provide the natural transport object for comparisons carried out at fixed $\Omega_\phi$. In that sense, $\Delta\psi_{\rm orb}$ is not proposed as a new raw observable, but as the convention-explicit scalar content of the usual nodal timing information.

The analysis is carried out for thin, slightly tilted circular rings outside ISCO in stationary and axisymmetric settings, with Boyer--Lindquist time used so that closed-form epicyclic expressions remain available. Within this setting the invariant packaging, the fixed-$\Omega_\phi$ transport calculus, and the leading analysis-level corrections can all be written explicitly. Beyond-Kerr structure is discussed through ACMC expansions, while a near-equatorial Johannsen--Psaltis proxy is used to illustrate deformation and spin dependence without tying the discussion to a single non-Kerr model \cite{Thorne1980,JohannsenPsaltis2011,Hansen1974}. The GRO J1655$-$40 benchmark is included to place the invariant quantity directly in the published observable domain.

\section{Orbital nodal phase and fixed-azimuthal-frequency transport}\label{sec:invariant}

We work in BL coordinates $(t,r,\theta,\phi)$ with $G=c=M=1$ and Kerr spin parameter $a$ \cite{Kerr1963,Carter1968}. On the equator $\theta=\pi/2$, we use $\Delta=r^{2}-2r+a^{2}$ and, with signs fixed,
\begin{equation}
g_{tt}=-(1-2/r),
\qquad
g_{t\phi}=-2a/r,
\qquad
g_{\phi\phi}=r^2+a^2+2a^2/r.
\end{equation}
These give the needed radial derivatives
\begin{equation}
\partial_r g_{tt}=-\frac{2}{r^{2}},
\qquad
\partial_r g_{t\phi}=\frac{2a}{r^{2}},
\qquad
\partial_r g_{\phi\phi}=2r-\frac{2a^{2}}{r^{2}}.
\end{equation}

The azimuthal BL-time frequency $\Omega$ of circular equatorial motion obeys the Bardeen--Press--Teukolsky condition \cite{BPT72}
\begin{subequations}\label{eq:BPT_block}
\begin{equation}
\partial_r g_{tt}
+
2\,\Omega\,\partial_r g_{t\phi}
+
\Omega^{2}\,\partial_r g_{\phi\phi}
=
0,
\label{eq:BPT_blocka}
\end{equation}
\begin{equation}
\Omega_\phi
=
\frac{1}{r^{3/2}+a},
\qquad
\partial_r\Omega_\phi
=
-\frac{3}{2}\,
\frac{r^{1/2}}{(r^{3/2}+a)^{2}}.
\label{eq:BPT_blockb}
\end{equation}
\end{subequations}
The epicyclic frequencies are taken in their standard BL-time forms \cite{BPT72}; we will only need the ratio
\begin{equation}
\frac{\Omega_\theta}{\Omega_\phi}
=
\sqrt{
1-\frac{4a}{r^{3/2}}+\frac{3a^{2}}{r^{2}}
}.
\label{eq:ratio_only}
\end{equation}

For a thin, slightly tilted ring let $\psi$ denote the nodal, or disc-normal, angle. Since $\dot\psi=\Omega_\phi-\Omega_\theta$ and $\dot\phi=\Omega_\phi$, the nodal phase accumulated per $2\pi$ in azimuth at fixed $r$ is
\begin{subequations}\label{eq:holonomy_block}
\begin{equation}
\frac{d\psi}{d\phi}
=
1-\frac{\Omega_\theta}{\Omega_\phi},
\label{eq:holonomy_blocka}
\end{equation}
\begin{equation}
\Delta\psi_{\rm orb}(r,a)
=
2\pi
\left(
1-\frac{\Omega_\theta}{\Omega_\phi}
\right).
\label{eq:holonomy_blockb}
\end{equation}
\end{subequations}
We refer to $\Delta\psi_{\rm orb}$ as the orbital nodal phase. In the present setting it is simply the invariant packaging of the usual nodal-frequency ratio on a circular ring. Thus the point is not to introduce a new raw observable, but to identify explicitly which combination of the standard timing frequencies remains unchanged under the benign pipeline maps considered here and can therefore be reported consistently across analyses. In any timing analysis that identifies $\Omega_\phi$ and $\Omega_\theta$, a convenient reporting pair is $(\Omega_\phi,\Delta\psi_{\rm orb})$.

Interpreting this quantity geometrically on the orbit cylinder links naturally to classical anholonomy ideas \cite{Hannay1985,Berry1984}. Introduce
\begin{equation}
\mathcal{A}
=
\left(
1-\Omega_\theta/\Omega_\phi
\right)
d\phi,
\end{equation}
so that along a circular ring, where $dr=0$, the ring holonomy is
\begin{equation}
\oint \mathcal{A}
=
2\pi
\left(
1-\Omega_\theta/\Omega_\phi
\right).
\end{equation}
Under a relabelling $\phi\mapsto\phi+\chi(r)$, the connection changes by an exact term plus a contribution proportional to $dr$, and therefore the ring holonomy on a circular ring remains unchanged. A second benign timing redefinition is time-azimuth mixing along a ring, $t'=\alpha t+\beta\phi$ with $\alpha+\beta\Omega_\phi>0$, which rescales both $\Omega_\phi$ and $\Omega_\theta$ by the same factor $(\alpha+\beta\Omega_\phi)^{-1}$ and therefore leaves Eq.~\eqref{eq:holonomy_block} unchanged. Monotonicity outside the ISCO follows by differentiating $f=\Omega_\theta/\Omega_\phi$ directly,
\begin{equation}
\partial_r
\left(
\frac{\Omega_\theta}{\Omega_\phi}
\right)
=
\frac{
3\,a\,r^{-3}
\left(
r^{1/2}-a
\right)
}{
\sqrt{
1-\dfrac{4a}{r^{3/2}}+\dfrac{3a^{2}}{r^{2}}
}
}.
\label{eq:monotone}
\end{equation}
For prograde spin $a>0$ and $r>\mathrm{ISCO}(a)$ this derivative is positive, while for $a=0$ it vanishes identically. Hence $\Delta\psi_{\rm orb}$ decreases with increasing $r$ for prograde Kerr spin and vanishes identically in the Schwarzschild limit. Monotonicity outside ISCO is consistent with the standard Kerr epicyclic hierarchy \cite{BPT72,Chandra}.

\begin{figure}[t]
\centering
\includegraphics[width=0.8\linewidth]{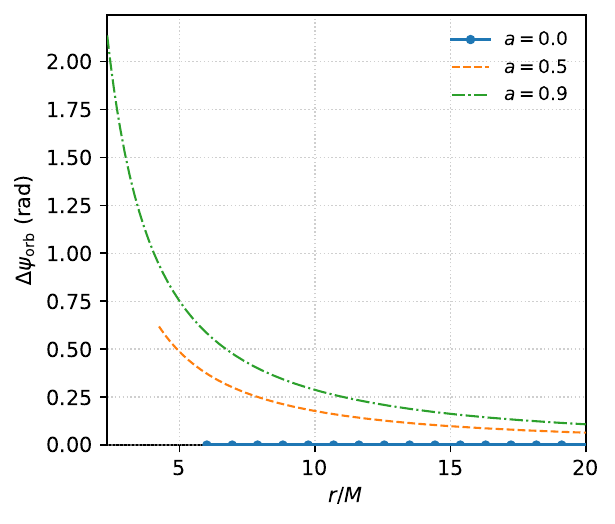}
\caption{Nodal phase per orbit $\Delta\psi_{\rm orb}$ versus $r/M$ for $a_\ast=\{0,0.5,0.9\}$; curves end at their ISCO.}
\label{fig:holonomy}
\end{figure}

Figure~\ref{fig:holonomy} illustrates this monotonicity, with $\Delta\psi_{\rm orb}$ decreasing with $r$ for prograde spin $a>0$ and steepening as the ISCO is approached, while in the Schwarzschild limit, $a=0$, it vanishes identically. At large radii the phase tends to zero as $\Delta\psi_{\rm orb}\sim 4\pi a\,r^{-3/2}+O(r^{-2})$, consistent with Eq.~\eqref{eq:ratio_only}.

To separate genuine multipolar physics from trivial radius drift, responses are taken at fixed $\Omega_\phi$. Let the metric be $g_{\mu\nu}(r,\theta;\epsilon)=g_{\mu\nu}^{(0)}(r,\theta)+\epsilon h_{\mu\nu}(r,\theta)+O(\epsilon^2)$, where $g_{\mu\nu}^{(0)}$ is the Kerr metric and $h_{\mu\nu}=\partial_\epsilon g_{\mu\nu}|_{\epsilon=0}$ is a stationary, axisymmetric first-order perturbation. Differentiating Eq.~\eqref{eq:BPT_block} at fixed $r$ gives
\begin{subequations}\label{eq:resp_block}
\begin{equation}
\partial_\epsilon \Omega_\phi\Big|_{r}
=
-\frac{
\partial_r h_{tt}
+
2\,\Omega_\phi\,\partial_r h_{t\phi}
+
\Omega_\phi^{2}\,\partial_r h_{\phi\phi}
}{
2\,\mathcal{D}
},
\label{eq:resp_blocka}
\end{equation}
\begin{equation}
\mathcal{D}
=
\sqrt{
(\partial_r g_{t\phi})^{2}
-
\partial_r g_{tt}\,\partial_r g_{\phi\phi}
}
=
\frac{2}{\sqrt{r}}.
\label{eq:resp_blockb}
\end{equation}
\end{subequations}
Imposing $d\Omega_\phi=0$ in the $(r,\epsilon)$ plane yields the transport rule
\begin{equation}
\partial_\epsilon r\Big|_{\Omega_\phi}
=
-\frac{
\partial_\epsilon \Omega_\phi\big|_{r}
}{
\partial_r \Omega_\phi
}.
\label{eq:transport_r}
\end{equation}
This transport requires $\partial_r\Omega_\phi\neq0$, which holds for circular equatorial orbits outside ISCO for prograde Kerr. For any scalar observable $X(r,\epsilon)$ this induces
\begin{equation}
\partial_\epsilon X\Big|_{\Omega_\phi}
=
\partial_\epsilon X\Big|_{r}
+
(\partial_r X)\,
\partial_\epsilon r\Big|_{\Omega_\phi}.
\end{equation}

Two ingredients enter repeatedly at fixed $\Omega_\phi$. The radial derivative of the vertical frequency follows from Eqs.~\eqref{eq:BPT_block} and \eqref{eq:ratio_only},
\begin{equation}
\partial_r \Omega_\theta^{2}
=
2\,\Omega_\phi\,(\partial_r \Omega_\phi)
\left(
1-\frac{4a}{r^{3/2}}+\frac{3a^{2}}{r^{2}}
\right)
+
\Omega_\phi^{2}
\left(
\frac{6a}{r^{5/2}}-\frac{6a^{2}}{r^{3}}
\right),
\label{eq:dOmth2}
\end{equation}
and the fixed-$\Omega_\phi$ sensitivity of the observable ratio may be summarized as
\begin{equation}
\Xi(r;a)
=
\left.
\partial_\epsilon
\left(
\frac{\Omega_\theta}{\Omega_\phi}
\right)
\right|_{\Omega_\phi}
=
\frac{\Omega_\theta}{\Omega_\phi}
\left[
\frac{1}{2}
\frac{
\partial_\epsilon \Omega_\theta^{2}\big|_{\Omega_\phi}
}{
\Omega_\theta^{2}
}
\right].
\label{eq:Xi_def}
\end{equation}
Thus $\delta(\Omega_\theta/\Omega_\phi)=\Xi\,\epsilon$ to first order while $\Omega_\phi$ is kept fixed. This isolates the response of the observable ratio under fixed-$\Omega_\phi$ transport from the radius adjustment implicitly required by an observer who indexes data by $\Omega_\phi$.

\section{Far-field scalings and observable consequences}\label{sec:conseq}

In ACMC coordinates the far-field $tt$ metric encodes the leading quadrupole deviation $\delta Q$ through the Legendre polynomial $P_2(\cos\theta)$ \cite{Thorne1980,Ryan1995}:
\begin{subequations}\label{eq:ACMC_block}
\begin{equation}
g_{tt}
=
-1+\frac{2}{r}
+
\frac{2\,\delta Q}{r^3}P_2(\cos\theta)
+
O(r^{-4}),
\label{eq:ACMC_blocka}
\end{equation}
\begin{equation}
\left.\partial_{\delta Q}g_{tt}\right|_{\theta=\pi/2}
=
-r^{-3},
\qquad
\left.\partial_r\partial_{\delta Q}g_{tt}\right|_{\theta=\pi/2}
=
3r^{-4},
\qquad
P_2(0)
=
-\frac{1}{2}.
\label{eq:ACMC_blockb}
\end{equation}
\end{subequations}
We write the mass quadrupole as $Q = Q_{\rm Kerr} + \delta Q$ with $Q_{\rm Kerr} = -a^2$ in the present $G=c=M=1$ units, so that $\delta Q$ measures deviations from Kerr in the normalization fixed by Eq.~\eqref{eq:ACMC_block}. Substituting the leading ACMC quadrupole term in Eq.~\eqref{eq:ACMC_block} into the fixed-$r$ response for $\Omega_\phi$, Eq.~\eqref{eq:resp_block}, with $\mathcal D=2/\sqrt r$ yields the leading far-field result
\begin{equation}
\partial_{\delta Q}\Omega_\phi\Big|_{r}
=
-\frac{3}{4}\,r^{-7/2}
+
O(r^{-9/2}).
\label{eq:phiACMC}
\end{equation}
Keeping $\Omega_\phi$ fixed then determines the corresponding leading induced radial shift from Eq.~\eqref{eq:transport_r} and $\partial_r\Omega_\phi$ in Eq.~\eqref{eq:BPT_block}:
\begin{equation}
\partial_{\delta Q} r\Big|_{\Omega_\phi}
=
-\frac{
\partial_{\delta Q}\Omega_\phi|_{r}
}{
\partial_r\Omega_\phi
}
=
-\frac{(r^{3/2}+a)^2}{2\,r^{4}}.
\label{eq:drACMC}
\end{equation}
Propagating the fixed-$\Omega_\phi$ shift into the vertical sector naturally separates the response into a transported Kerr radial contribution and a direct perturbative contribution at fixed radius. Using Eqs.~\eqref{eq:dOmth2} and \eqref{eq:drACMC}, the leading fixed-$\Omega_\phi$ ratio response may therefore be organised as
\begin{subequations}\label{eq:XiACMCsplit}
\begin{equation}
\Xi_{\mathrm{ACMC}}(r;a)
=
\Xi_{\mathrm{dir}}(r;a)
+
\Xi_{\mathrm{tr}}(r;a),
\label{eq:XiACMCsplita}
\end{equation}
\begin{equation}
\Xi_{\mathrm{dir}}(r;a)
=
\frac{\Omega_\theta}{\Omega_\phi}
\left[
\frac{1}{2}
\frac{
\partial_{\delta Q}\Omega_\theta^{2}\big|_{r}
}{
\Omega_\theta^{2}
}
\right],
\label{eq:XiACMCsplitb}
\end{equation}
\begin{equation}
\Xi_{\mathrm{tr}}(r;a)
=
\frac{\Omega_\theta}{\Omega_\phi}
\left[
\frac{1}{2}
\frac{
\left(\partial_r\Omega_\theta^{2}\right)
\partial_{\delta Q}r\big|_{\Omega_\phi}
}{
\Omega_\theta^{2}
}
\right].
\label{eq:XiACMCsplitc}
\end{equation}
\end{subequations}
This decomposition identifies how far-field quadrupolar and higher-multipolar timing responses enter the observable ratio at fixed $\Omega_\phi$.

Equatorial parity implies that, near Kerr, even multipoles dominate the beyond-Kerr correction to the nodal holonomy. The fixed-$\Omega_\phi$ calculus therefore provides a direct way to parameterise departures from the Kerr baseline in timing language rather than in coordinate coefficients alone. In an ACMC far field, Eqs.~\eqref{eq:phiACMC}, \eqref{eq:drACMC}, and \eqref{eq:XiACMCsplit} identify how far-field quadrupolar and higher-multipolar timing responses enter the observable ratio. Thus $\Delta\psi_{\rm orb}$ furnishes the timing-side Kerr baseline and the fixed-$\Omega_\phi$ framework relative to which strong-gravity deviations should be compared.

Because $\Delta\psi_{\rm orb}$ is scalar and, for prograde Kerr spin, monotone with $r$ outside ISCO, it also serves as a pipeline-robust ordering variable across source states and instruments. This is particularly useful when combining timing with reflection spectroscopy in joint constraints, because the invariant nodal phase provides a Kerr baseline map in the $(r/M,a)$ plane against which deformation-induced and source-level shifts may be compared \cite{BambiReflection2017}, suggesting complementarity in multi-probe analyses.

\begin{figure}[t]
\centering
\includegraphics[width=0.8\linewidth]{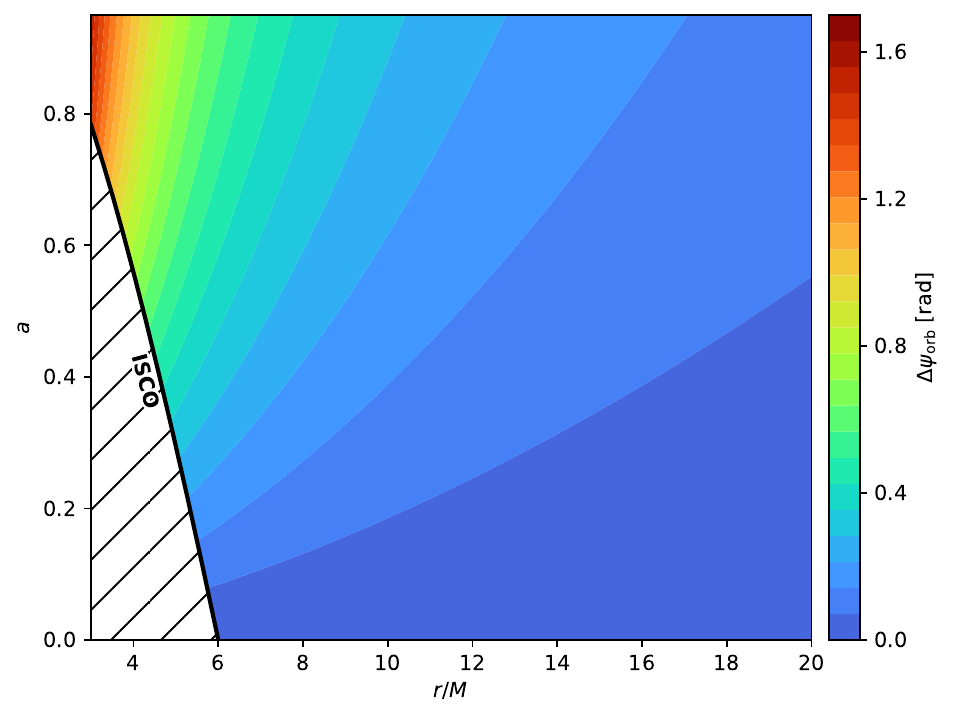}
\caption{Orbital nodal phase $\Delta\psi_{\rm orb}(r,a)$; filled bands show $\Delta\psi_{\rm orb}$ in radians, with ISCO locus, solid, and ISCO-forbidden region, hatched.}
\label{fig:xi}
\end{figure}

Figure~\ref{fig:xi} visualizes the orbital nodal phase $\Delta\psi_{\rm orb}(r,a)$ across the $(r/M,a)$ plane. Outside the ISCO, the phase decreases with radius for prograde spin and increases with spin at fixed radius. The ISCO boundary removes the inner wedge of parameter space.

A near-equatorial Johannsen--Psaltis proxy provides a convenient language for discussing deformation and spin dependence in this transport setting without tying the discussion to a single non-Kerr model \cite{JohannsenPsaltis2011}.

The preceding Kerr and near-Kerr calculations define the geodesic baseline for $\Delta\psi_{\rm orb}$ and its metric-side response at fixed $\Omega_\phi$. In more realistic accretion flows, additional shifts will in general arise from finite thickness, magnetic stresses, truncation, dissipation, and emission geometry, so the quantity advocated here should be viewed as a consistency diagnostic rather than a stand-alone identifier of the metric \cite{FragileLiska2024,Liska2023PhaseLag,Musoke2023Tearing,Bollimpalli2023TypeC,Bollimpalli2024Truncated,Moriyama2025RZ,RezzollaZhidenko2014}. In particular, this makes $\Delta\psi_{\rm orb}$ a natural comparison quantity for synthetic timing from tilted or precessing GRMHD flows, because it separates the Kerr geodesic baseline, the metric-side shift away from Kerr, and the residual matter-side modification introduced by the flow and emission model. This is precisely the structure required for a meaningful timing-side consistency test in which one asks whether a measured or synthetic signal can still be reconciled with the Kerr baseline without invoking implausibly large flow-side corrections.

Consider a thin, slightly tilted ring whose radius executes a small coherent breathing at the radial epicyclic frequency, as in thin-disc oscillation systematics \cite{Kato2001,Torok2005}, $r(t)=r_0+A\cos(\Omega_r t)$ with $A\ll r_0$. Writing $N(r)=\Omega_\theta(r)$, $D(r)=\Omega_\phi(r)$, and $R(r)=N(r)/D(r)$, the relevant average for the accumulated nodal phase per azimuthal orbit is the azimuth-weighted average, not a plain time average. Expanding this average about $r_0$ through second order, with $\langle \delta r\rangle=0$ and $\langle \delta r^2\rangle=A^2/2$, gives
\begin{equation}
\left\langle
\frac{\Omega_\theta}{\Omega_\phi}
\right\rangle_{\rm orb}
=
\frac{N_0}{D_0}
+
\frac{A^2}{2}\,C_2(r_0,a)
+
O(A^4),
\label{eq:avgRatio_compact}
\end{equation}
where $N_0=\Omega_\theta(r_0)$, $D_0=\Omega_\phi(r_0)$, $R(r)=\Omega_\theta(r)/\Omega_\phi(r)$, and
\begin{equation}
C_2(r_0,a)
=
\frac{1}{2}
\left[
\left.
\frac{d^2R}{dr^2}
\right|_{r=r_0}
+
2
\left.
\frac{dR}{dr}
\right|_{r=r_0}
\left.
\frac{d\ln\Omega_\phi}{dr}
\right|_{r=r_0}
\right].
\label{eq:C2_definition}
\end{equation}
This is the exact second-order coefficient obtained from the Kerr expressions for $\Omega_\phi$ and $\Omega_\theta$. Using the Kerr expressions for $\Omega_\phi$ and $\Omega_\theta$, two immediate properties follow: in the Schwarzschild limit $a\to0$ one has $\Omega_\theta\equiv\Omega_\phi$ and hence $C_2\to0$; in the far field $r_0\gg M$, a straightforward large-$r_0$ expansion yields
\begin{equation}
C_2(r_0,a)
=
-\frac{33}{4}\,\frac{a}{r_0^{7/2}}
+
9\,\frac{a^2}{r_0^{4}}
+
O(r_0^{-9/2}),
\label{eq:C2asym_compact}
\end{equation}
so for prograde spin the leading term is negative. The mean per-orbit nodal phase therefore acquires a positive second-order bias proportional to the variance of the breathing,
\begin{equation}
\delta
\left[
\Delta\psi_{\rm orb}
\right]
=
-\pi\,C_2(r_0,a)\,A^2
+
O(A^4).
\label{eq:breathBias_compact}
\end{equation}
The sign and scaling are thus fixed by the Kerr background alone, with the bias vanishing in Schwarzschild, growing with spin, and decaying with radius roughly as $r_0^{-7/2}$ at leading order.

Slow changes in two external controls can be organized by the fixed-$\Omega_\phi$ transport introduced earlier. Let $\varepsilon^A$ denote two slow parameters and define the connection coefficients
\begin{equation}
\Gamma_A
=
\frac{\partial_A\Omega_\phi|_{r}}{\partial_r\Omega_\phi},
\end{equation}
so that derivatives at fixed $\Omega_\phi$ act as
\begin{equation}
\partial_A\Big|_{\Omega_\phi}
=
\partial_A
-
\Gamma_A\,\partial_r.
\end{equation}
Writing $\omega(r,\varepsilon)=\Omega_\phi(r,\varepsilon)$, this gives $\Gamma_A=\omega_A/\omega_r$. The associated curvature combination is
\begin{equation}
F_{AB}
=
\partial_A\Gamma_B
-
\partial_B\Gamma_A
-
\Gamma_A\,\partial_r\Gamma_B
+
\Gamma_B\,\partial_r\Gamma_A.
\end{equation}
For the exact fixed-$\Omega_\phi$ transport defined by a single scalar constraint, this curvature vanishes identically:
\begin{equation}
F_{AB}
=
0,
\qquad
\big[
\partial_A|_{\Omega_\phi},
\partial_B|_{\Omega_\phi}
\big]\,R
=
0.
\label{eq:comm_holonomy}
\end{equation}
The corresponding transported offset after a slow, closed loop $\mathcal C$ in control space is therefore
\begin{equation}
\Delta_{\mathcal C}R
=
0
\label{eq:loop_anholonomy}
\end{equation}
for any oriented surface $\mathcal S$ with boundary $\partial\mathcal S=\mathcal C$.

This null result is useful because it separates exact coordinate transport from genuine source dynamics. A nonzero loop memory in timing data would require physics beyond the instantaneous geodesic map, such as dissipative lag, hysteresis, nonunique flow branches, finite response time, or additional internal accretion-flow variables. Such effects can still be compared against the fixed-$\Omega_\phi$ Kerr baseline, but they are not generated by the scalar constraint $\Omega_\phi=\mathrm{constant}$ itself.

We restrict the quantitative benchmark to GRO J1655$-$40 because it provides a clean internally consistent triplet set under a single published identification convention. As a limited published-data benchmark, we restrict attention to the canonical GRO J1655$-$40 triplet measurements. Following Ref.~\cite{Motta2014GRO}, the upper high-frequency quasi-periodic oscillation, HFQPO, is taken as the orbital-frequency anchor, and for the GRO B1 row the same averaging procedure is used. We do not use GRS 1915+105 quantitatively here, so as to avoid mixing source-dependent identification choices within what is intended only as a single-source benchmark \cite{MottaBelloni2024}.

\begin{table}[t]
\centering
\caption{Published GRO J1655$-$40 timing cases and reconstructed values of $\Delta\psi_{\rm orb}$, compiled from Ref.~\cite{Motta2014GRO}.}
\label{tab:published_cases_raw}
\small
\begin{tabular}{@{}llcccc@{}}
\toprule
Source
&
Case
&
$\nu_{\rm nod}$ [Hz]
&
$\nu_{\rm HF}$ [Hz]
&
${\Delta\psi}_{\rm orb}/(2\pi)$
&
${\Delta\psi}_{\rm orb}$ [rad]
\\
\midrule
GRO J1655$-$40
&
B1 average
&
$17.3\pm0.1$
&
$441\pm2$
&
$0.03923\pm0.00029$
&
$0.2465\pm0.0018$
\\
GRO J1655$-$40
&
10255-01-09-00
&
$18.3\pm0.1$
&
$451\pm5.5$
&
$0.04058\pm0.00054$
&
$0.2549\pm0.0034$
\\
GRO J1655$-$40
&
10255-01-10-00
&
$18.1\pm0.1$
&
$446\pm4$
&
$0.04058\pm0.00043$
&
$0.2550\pm0.0027$
\\
\bottomrule
\end{tabular}
\end{table}

The GRO B1 row follows the averaging procedure adopted in Ref.~\cite{Motta2014GRO}; the upper HFQPO is used as the orbital-frequency anchor. For the 10255-01-09-00 case, the published HFQPO uncertainty is asymmetric, $451^{+6}_{-5}$ Hz in Ref.~\cite{Motta2014GRO}; it has been symmetrised to $451\pm5.5$ Hz, taking the mean of the two sides, for the purpose of simple error propagation.

For the three GRO J1655$-$40 benchmark cases, the reconstructed values are ${\Delta\psi}_{\rm orb}=0.2465\pm0.0018,\ 0.2549\pm0.0034,\ 0.2550\pm0.0027\ {\rm rad}$. These values occupy a narrow range around $0.25$ rad under the adopted RPM identification. This benchmark places $\Delta\psi_{\rm orb}$ directly in the published observable domain under a standard RPM identification. Once the orbital-frequency anchor is specified, the invariant quantity can be reconstructed from standard timing products and quoted with propagated uncertainty. It is therefore immediately usable as a common comparison quantity for observational timing papers, strong-gravity transport-based comparisons, and synthetic timing studies.

\begin{figure}[t]
\centering
\includegraphics[width=0.80\linewidth]{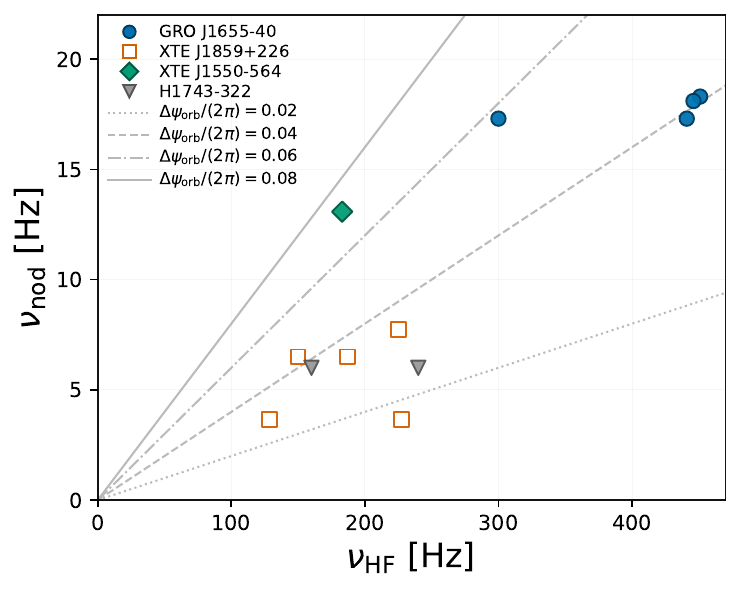}
\caption{Published timing cases in the raw observable plane $(\nu_{\rm HF},\nu_{\rm nod})$. The three GRO J1655$-$40 cases from Table~\ref{tab:published_cases_raw} are the only cases used quantitatively in the text. Additional literature points are shown only to illustrate the geometry of constant-$\Delta\psi_{\rm orb}$ rays in this plane. Grey rays indicate constant values of the observational proxy $\Delta\psi_{\rm orb}/(2\pi)=\nu_{\rm nod}/\nu_{\rm HF}$.}
\label{fig:obscheck}
\end{figure}

Figure~\ref{fig:obscheck} places the GRO J1655$-$40 benchmark cases in the raw observable plane $(\nu_{\rm HF},\nu_{\rm nod})$ and overlays constant-$\Delta\psi_{\rm orb}$ rays for geometric context. Additional published measurements are included to display how this invariant quantity is organised in the observable plane. Quantitative discussion in the text remains anchored to the GRO J1655$-$40 benchmark.

For observational use, once the orbital-frequency anchor and component identification have been specified, the propagated uncertainty in the reconstructed $\Delta\psi_{\rm orb}$ under the same RPM convention used above and with negligible covariance between the measured frequencies is
\begin{equation}
\sigma_{\Delta\psi_{\rm orb}}
=
2\pi
\sqrt{
\frac{\sigma_{\nu_{\rm nod}}^2}{\nu_{\rm HF}^2}
+
\frac{\nu_{\rm nod}^2\,\sigma_{\nu_{\rm HF}}^2}{\nu_{\rm HF}^4}
}.
\end{equation}
A convenient reporting convention is therefore to list $\nu_{\rm HF}$, $\nu_{\rm nod}$, the reconstructed $\Delta\psi_{\rm orb}$, its propagated uncertainty, and the adopted orbital-frequency anchor and identification convention. In that sense, the minimal observer-facing summary advocated here is not just the measured frequency pair, but the pair together with the reconstructed invariant quantity and the convention used to define it.

\section{Discussion}\label{sec:disc}

Within the thin-ring, slightly tilted, stationary setting considered here, the central result is the identification of the pair $(\Omega_\phi,\Delta\psi_{\rm orb})$, with $\Delta\psi_{\rm orb}$ the nodal phase accumulated per orbital cycle and invariant under the admissible time and azimuth redefinitions of Sec.~\ref{sec:invariant}. For a single idealised orbit this quantity is algebraically equivalent to the usual nodal-frequency ratio up to the factor $2\pi$. The point is therefore not to introduce a new raw observable, but to identify explicitly the convention-explicit scalar that remains unchanged under the benign remappings relevant to timing pipelines and that is therefore safe to compare across analyses. The geometric formulation becomes necessary at the level of structure rather than algebra. It provides the direct invariance proof, identifies the natural transport object on the orbit cylinder, and clarifies the transport viewpoint relevant when comparisons are made at fixed $\Omega_\phi$ rather than at fixed coordinate radius. Without that step, one may still manipulate the same ratio numerically, but the convention-independent content and the corresponding transport structure are left implicit.

The GRO J1655$-$40 benchmark places the construction directly in the observational plane. Once the orbital-frequency anchor and component identification are stated, $\Delta\psi_{\rm orb}$ can be reconstructed from standard published QPO frequencies and quoted with propagated uncertainty. The same quantity can therefore be carried consistently between observational timing products, strong-gravity transport-based comparisons, and synthetic timing or GRMHD-based studies.

Two analysis-level results follow naturally from the framework. Coherent radial breathing at $\Omega_r$ shifts the mean nodal phase at second order, while exact slow two-parameter evolution gives no intrinsic geometric offset under fixed-$\Omega_\phi$ transport. Their source-specific observability should be assessed within end-to-end spectral-timing analyses that include broad-band noise, finite coherence, frequency drift, non-stationarity, and component-identification systematics \cite{Nathan2022,Zhu2024,Yu2024,Bollemeijer2025}.

The present formulation is deliberately analytic, being built on infinitesimally thin, slightly tilted circular rings in stationary and axisymmetric settings, with Boyer--Lindquist time used to preserve closed-form epicyclic expressions. Extending the present transport calculus to thicker flows, radiative models, and specific beyond-Kerr families is the natural next step.

\section*{Acknowledgements}

The author thanks A.\ S.\ Turan for helpful discussions and the anonymous referee for constructive comments that improved the manuscript. Parts of this work were carried out while the author was at Ludwig Maximilian University of Munich and the University of Warwick.

\bibliographystyle{unsrtnat}
\bibliography{bib}

\end{document}